\documentclass {article}
\usepackage{amsfonts}
\DeclareSymbolFont{AMSb}{U}{msb}{m}{n}
\DeclareSymbolFontAlphabet{\mathbb}{AMSb}
\newcommand{\complex}{\kern.1em{\raise.47ex\hbox{
            $\scriptscriptstyle |$}}\kern-.40em{\rm C}}
\newcommand{\ket}[1]{\left\vert #1 \right\rangle}
\newcommand{\bra}[1]{\left\langle #1 \right\vert}
\newcommand{\kernel}[3]{\left\langle #1\left\vert #2\right\vert#3 \right\rangle}

\input epsf
\title{Topological quantum numbers in the Hall effect}
\begin{document}

\author{J.~E.~Avron$^1$, D. Osadchy$^1$ and R. Seiler$^2$
\\ $^1$ Department of Physics, Technion, 32000 Haifa, Israel
\\ $^2$ Department of Mathematics, TU Berlin, Berlin, Germany }

\maketitle


\input epsf
\begin{abstract}

Topological quantum numbers account for the precise quantization that occurs in the integer Hall effect.  In this  theory, Kubo's formula for the conductance acquires a  topological interpretation in terms of Chern numbers and their non-commutative analog, the Fredholm Indices.
\end{abstract}

\section{The Hall effect}

The story of the  Hall effect begins with a mistake made by James
Clerk Maxwell, (1831-1879). In the first edition of his book
\emph{``Treatise on Electricity and Magnetism''}, which appeared
in 1873, Maxwell discussed the deflection of a current carrying
wire by a magnetic field. Maxwell then says: \emph{It must be
carefully remembered that the mechanical force which urges a
conductor \dots acts, not on the electric current, but on the
conductor which carries it}.  If the reader is puzzled that is OK,
he should be.

In  1878 Edwin H.~Hall, a student at Johns Hopkins University, was
reading  Maxwell for a class by Henry A. Rowland. Hall was puzzled
by this passage and approached  Rowland. Rowland told him that  \cite{ref:hall}
\emph{...he doubted the truth of Maxwell statement and had
sometimes before made a hasty experiment \dots though without
success.}  Hall
made a fresh start, and tried to measure the magnetoresistence---a hard
experiment. This experiment failed too and Maxwell appeared to be safe. Hall then decided
to repeat the experiments made by Rowland, and following a
suggestion of his advisor, replaced the original metal bar with
a thin gold leaf and found that the magnetic field deflected the
galvanometer needle.  This  earned Hall a position at Harvard.

Maxwell died in the year that Hall's paper came out. In the second
edition of Maxwell's book, which appeared posthumously in 1881,
there is a  polite footnote by the editor saying: \emph{Mr. Hall has
discovered that a steady magnetic field does slightly alter the
distribution of currents in most conductors so that the statement
in brackets must be regarded as only approximately true.} It
turned out that the magnitude, and even the sign of the Hall
voltage depends on the conductor. This made the Hall effect an
important diagnostic tool. Maxwell, even in error, inspired a
remarkable research direction.

A hundred years later, the Hall effect was revived as a source of wonderful physics.
In 1980 Klaus von Klitzing discovered that two dimensional electron gas,
at very low temperatures and strong magnetic fields, displays a remarkable quantization of the Hall conductance. Namely, the graph of the Hall conductance as function of the magnetic field, is a staircase function, where the value of the Hall conductance at the plateaus is, to great accuracy, an integer multiple of
$e^2/h=1/(25812.807572\,\Omega)$. This discovery led to superior standards of resistance and 
von-Klitzing was awarded the Nobel prize in 1985 for his discovery.

The precision of the quantization in the Hall effect is remarkable in that it takes place in systems that
are imprecisely characterized on the microscopic scale: Different
samples have different distributions of impurities, different
geometry and different concentrations of electrons. Nevertheless,
whenever their Hall conductances are quantized, the quantized
values mutually agree with great precision.

The quantum Hall effect may also be interpreted as a measurement of the fine structure constant. The precision is slightly inferior to the determination that follows from the measurement of the anomalous magnetic moment of the electron: The Hall effect gives $137.0360\,0300(270)$ for the inverse of the fine structure constant while the anomalous magnetic moment gives $137.0359\,9976(50)$. Interestingly, the latter will need to be revised by 6 ppb  due to an error in the computation of 18 Feynmann diagrams \cite{ref:Kinoshita}.

Among the deep theoretical developments spawned  by this discovery
was the recognition that the Hall conductance has 
topological significance \cite{ref:tknn,ref:bell}. In
particular, its precise quantization can be understood in terms of
topological invariants known as Chern numbers and their
non-commutative analog: Fredholm Indices. These topological
invariants are our theme.

\section{Laughlin argument}\label{sec:average}

In a 1981 Robert Laughlin \cite{ref:laughlin} put forward an
argument for the  quantization of the Hall conductance. This
argument played  a seminal role in the development of the theory of the Integer Hall effect and it deserves to be re-examined a quarter of a century later. As we shall see, Laughlin argument goes most of the way in explaining the quantization, short of one step: The step of topological quantum numbers.

\begin{figure}[h]
\hskip .5 in \epsfxsize=4.in \epsfbox{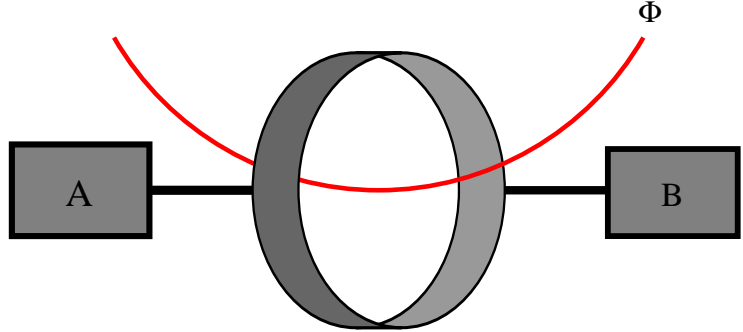} \caption{The Hall
effect can be interpreted as a quantum pump. Increasing the flux
$\Phi$ that threads the two dimensional ribbon of a 2D electron
gas, by a unit of quantum flux, transfers charges from one
reservoir to the other. Under appropriate conditions the average
transferred charge is quantized.} \label{fig:corbino}
\end{figure}

Laughlin considers a 2D electron gas  confined to the ribbon
shown in  Fig.~\ref{fig:corbino}. In this geometry, the Hall effect may be interpreted as a pump  that transfers charges from the reservoir $A$ to the reservoir $B$. The pump is driven by magnetic flux tube in the figure carrying a time dependent flux $\Phi$. A cycle of the pump corresponds to an increase of $\Phi$ by a unit of quantum flux $\Phi_0=hc/e$.  The Hall conductance, measured in the quantum  unit of conductance, $e^2/h$, is then the number of electrons transported between the reservoirs in a cycle of the pump.  Laughlin concludes the argument saying:{\em  \dots  by
gauge invariance, adding $\Phi_0$ maps the system back to itself,
\dots [resulting in] the transfer of n electrons}. The
quantization of the Hall conductance is  then implied.

That the number of particles transferred to the $A$ reservoir is an integer follows from the a basic tenet of quantum mechanics which guarantees that each measurement of the number of particles in $A$ always gives an integer.  However, there is no general a-priori reason why each and every cycle of the pump should transfer {\em the same number} of particles. In a quantum theory this number may  be a fluctuating integer.

More precisely, the Hall conductance is the {\em average}  number of particles transferred in a cycle of the pump. To complete the Laughlin argument one therefore needs to understand when and why quantum  {\em averages} are quantized. This is where topological quantum numbers come into play. As we shall explain, {\em{topological quantum numbers quantize averages.}}

An alternate way of formulating the issue at stake is as follows: The Hall conductance  is determined by Kubo's formula. Being a formula for a quantum average it takes the form of a quantum expectation value. Why is it quantized? As we shall explain, Kubo's formula can be identified with known topological invariants: Chern numbers and their non-commutative analogs, Fredholm indices.

\section{The geometry of linear response}

At about the same time that these developments took place in the
quantum Hall effect, another theoretical development unfolded.
In 1981 Michael V. Berry \cite{ref:berry} discovered that the phase
accumulated by the wave function undergoing adiabatic evolution
can be decomposed into two pieces---a dynamical part that carries
information on the energy of the system, and a part, Berry's phase,
that carries geometric information on the wave function. To explain the geometric significance of the Berry's phase
it is instructive to take a step back and reconsider the geometry of surfaces.

In 1917 Tulio Levi-Civita developed a fresh perspective on the geometry
of surfaces by focusing on parallel transport as the primary object. The face of the 
earth is an example of a curved surface---a sphere---and  the Foucault 
pendulum realizes the notion of parallel transport on it. 

Consider a Foucault pendulum at latitude $\theta$. The plane of
the pendulum defines a direction on earth. As the earth rotates, 
the angle between the direction of the pendulum and the latitude changes 
as if the pendulum was parallel transported along a 
circle of constant latitude. Once earth completed a cycle the  
pendulum  does not point in the same direction as it did initially. Rather, there is an
angular mismatch of  $2\pi\,\sin\theta$. This mismatch is
geometric: It is independent of the angular velocity of earth
around its axis (so long as it is small compared with the natural
frequency of the pendulum). More importantly, it is a hallmark of curvature. 
So, while the conventional point of
view is to regard the Foucault pendulum as a demonstration that 
earth is  rotating, from the geometric point of view it 
may also be interpreted as a demonstration that earth is a sphere.

Viewing curvature as the mismatch of parallel transport allows the
extension of the notion of curvature beyond geometry to quantum mechanics.
The wave function, a vector in the Hilbert space, plays a role
analogous to the direction determined by the plane of the Foucault
pendulum. A full rotation of the earth is an analog of taking the Hall
pump through a complete cycle.  The Berry's phase which is accumulated in a cycle of the pump is the analog of the  angular mismatch accumulated by the pendulum in a day.

To complete the analogy and introduce the notion of adiabatic curvature one 
needs an analog of longitude. This is  
$\theta$, the phase associated with a gauge transformation 
twisting the two reservoirs. The corresponding eigenfunctions, $\ket{\psi}$, then 
depend on two parameters $\Phi$ and $\theta$ and Berry's phase accumulated 
by traversing a small loop is proportional to the area of the loop. The adiabatic curvature,
\begin{equation}\label{eq:berry-curv}
K = 2\, Im \bra{\partial_\Phi\psi}\partial_\theta\psi\rangle .
\end{equation}
is the proportionality factor.

The significance of the adiabatic curvature to transport comes from the fact that it can be identified  with conductance as determined by Kubo's formula. 
To see this note that the quantum mechanical current operator is 
$I=c\, \partial_\theta H$. In the adiabatic regime where $\dot\Phi$ 
is small one finds as a consequence of the Schr\"odinger equation that the expectation value for the current is: 
\begin{equation}\label{eq:q-transport}
\kernel{\psi}{I}{\psi}=c\,\partial_\theta E+\hbar\,c\,K\,\dot\Phi
\end{equation}
In the time-independent case, $\dot\Phi=0$, this  reduces to the more familiar Feynman-Hellmann theorem. Eq.~(\ref{eq:q-transport}) gives a linear relation between the current and the driving emf,  $\frac 1 c\, \dot\Phi$, generated by the flux tube. The conductance is therefore $\hbar c^2\, K$.  This expression is equivalent to Kubo's formula for the conductance in linear response theory.

\section{Chern numbers}

Ludwig Boltzmann is reputed to have said that elegance is for
tailors. The geometric interpretation of linear
response is clearly elegant. Is there more to it than just elegance?
There is: Geometry  links linear response with topological invariants.

Geometry and topology are, of course, intimately related. Let us
start by recalling this relation in its most intuitive setting:
The theory of surfaces.

\begin{figure}[h]
\hskip 1.4 in\epsfxsize=2.2 in \epsfbox{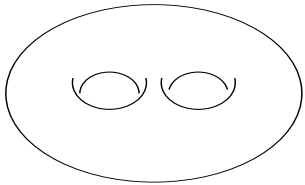}
\caption{A surface with two handles, $g=2$.} \label{fig:riemann}
\end{figure}

A remarkable relation between
geometry and topology is the formula by Gauss and Bonnet:
\begin{equation}\label{eq:gauss}
\frac 1 {2\pi} \int_S \, dA\, K =2 (1-g)
\end{equation}
$S$ is a  surface, without a boundary, like  the two handled torus in Fig.~\ref{fig:riemann}. $K$ is the (Gaussian) curvature of the surface
(i.e. $K^{-1}$ is the product of the two radii of curvatures) and $dA$ the area element. The data on the left are purely geometric. $g$ on the
right, the number of handles ($g=2$ in the
figure), is purely topological. In
particular, a deformation of the surface does not change the
number of handles. This then implies that the curvature of a
surface can not be deformed at will: Its integral is constrained by topology.

The Gauss-Bonnet formula has a generalization, known as
Gauss-Bonnet-Chern formula, which goes beyond the theory of
surfaces. The Chern number is the analog of $2(1-g)$ in Eq.~(\ref{eq:gauss}).  The integral of the adiabatic curvature over parameter space $S$ is quantized provided $S$ is compact and has no boundary.

\begin{figure}[h]
\hskip 1.4 in\epsfxsize=1.6 in \epsfbox{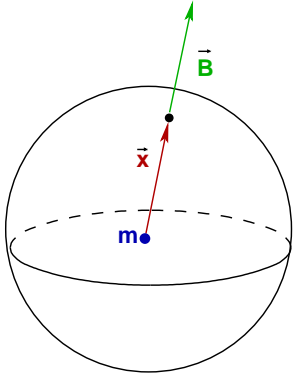}
\caption{The magnetic field of a monopole $\vec B$ is radial. The spin eigenstates at $\vec x$ are parallel and anti-parallel to $\vec B$. The phase of the spin wave function is forced to have a point of singularity somewhere on the sphere.} \label{fig:monopole}
\end{figure}

The simplest quantum mechanical example of a Chern number is a heavy spin $1/2$ particle on a spherical shell with a magnetic monopole at the center \cite{ref:berry}, see  Fig.\ref{fig:monopole}. The Hamiltonian is $\vec\sigma\cdot\vec B$ where $\vec\sigma $ is the triplet of Pauli matrices and $\vec B$, the magnetic field, is pointing in the radial direction. 

The ground state
singles out a spinor in  the two (complex) dimensional Hilbert
space.  Now, even though the Hamiltonian is a
continuous function of the position of the particle one can not demand that the ground state spinor, $\vert \chi(\vec B)\rangle $, be both
normalized  and continuous  everywhere on the sphere. This is an analog of the geometric fact that a vector field tangent to the sphere can not be both continuous and normalized.  At best, one may choose the ground state spinor to be normalized and
continuous away from a single point, say the south pole. We denote this choice by $\vert \chi_s(\vec B)\rangle$. Similarly, $\vert \chi_n(\vec B)\rangle$, is the spinor that is normalized and continuous away from the north pole. Away from the
poles the two choices of the ground state must agree up to a
phase, namely,
\begin{equation}
\vert \chi_n(\vec B)\rangle=e^{i\gamma(\vec B)}\vert \chi_s(\vec B)\rangle
\end{equation}
As $\vec B$ goes around a latitude, the phase winds.  For the ground
state of spin $1/2$ it winds once. This  obstructs extending
$\vert \chi_s(\vec B)\rangle$ to the south pole since
$\vert \chi_n(\vec B)\rangle$ is continuous there. The
obstruction is a phase aspect of the wave function. There
is no obstruction to choosing the projection on the ground state to be  continuous everywhere on the sphere.

The relation between Chern numbers and the winding of phase $\gamma$ explains why Chern numbers are topological.

\begin{figure}[h]
\center{ \epsfxsize=4.7in \epsfbox{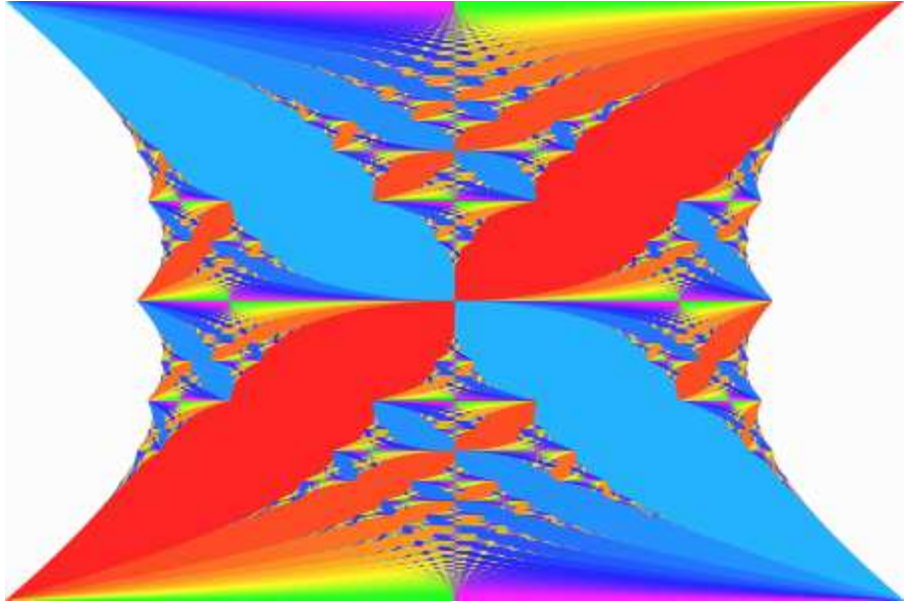} }
\caption{The phase diagram of  the tight-binding Hofstadter model. The colors represent different quantum Hall phases labelled by Chern numbers.
The horizontal axis is chemical potential which fixes the density of the electrons. The vertical axis
is the magnetic flux through the unit cell measured in units of quantum flux. 
The phase diagram is periodic in the vertical direction with a period of one quantum flux. It is also anti-symmetric under reflection in a horizontal and vertical lines. } \label{hb-tb}
\end{figure}

\section{Chern numbers in the Hall effect}

A simple yet often rather accurate model for 2D electron gas is 
the Landau Hamiltonian. In this model the Chern numbers count 
the filled Landau levels; so its success made the art of calculating Chern
numbers dispensable.
To display the glory of Chern numbers one needs to  consider situations which can not be approximated in terms of the Landau model. This is the case for  periodic 
\cite{ref:tknn} and multiply connected systems \cite{ref:stone}.

The tight-binding analog of the Landau model is known as the Hofstadter model. It has a plethora of Chern numbers with intricate dependence on the electron density and the magnetic field. The model is formally
defined by the one particle Hamiltonian
\begin{equation}\label{eq:harper}
H= U+U^*+V+V^*, \quad UVU^*V^*=e^{2i\pi\phi}
\end{equation}
where $U$  and $V$ are the operators of translation.
$\phi=\Phi/\Phi_0$ where  $\Phi$ is now the magnetic flux through unit cell.  Remarkably, precisely the same  model also describes  the splitting of the lowest Landau level by a super-lattice potential except that in this case $\phi=\Phi_0/\Phi$.

When the Fermi energy lies in a gap and $\phi$ is rational the Brillouin zone plays the role of $S$ in Eq.~(\ref{eq:gauss}) and  the Hall conductance at zero temperature can be identified with a Chern number.  The Chern numbers of the Hofstadter model can be determined as the solution of  certain  Diophantine equations \cite{ref:tknn}.
\begin{figure}[h]
  \centering
 \epsfxsize=3.5 in  \epsfbox{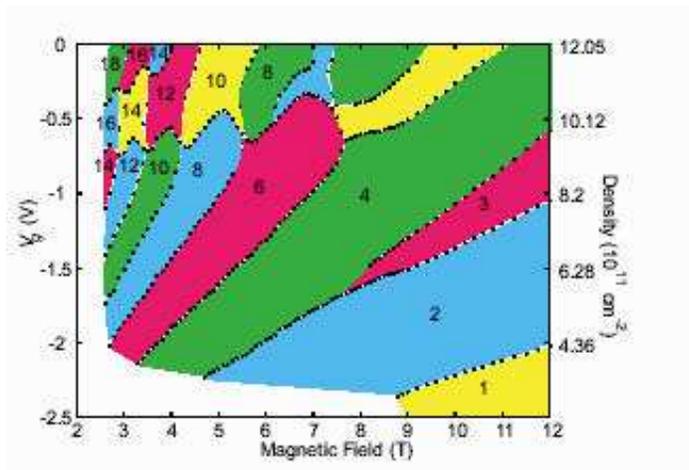}
  \caption{The quantum phase diagram of a bi-layer 2D electron gas taken from the home page of Jiang \cite{ref:jiang}. The similarity with the tips of Fig.~\ref{hb-tb}, as well as the important differences, are noteworthy. Note that the axis are flipped relative to the axis in Fig.~\ref{hb-tb}. }\label{fig:jiang}
\end{figure}

The $T=0$ quantum phase diagram of the tight-binding Hofstadter model is shown in Fig.~\ref{hb-tb}.  It is a fractal phase diagram where the colors represent distinct quantum phases, labelled by  Chern numbers. Warm colors represent positive Chern numbers, cold colors represent negative ones and white is zero.  Experiments, such as those shown in Fig.~\ref{fig:jiang}, explore a thin horizontal sliver of the diagram near $\Phi=0$, because available magnetic fields
have tiny flux through the unit cell of natural crystals. 

\begin{figure}
\center{ \epsfxsize=4.7 in \epsfbox{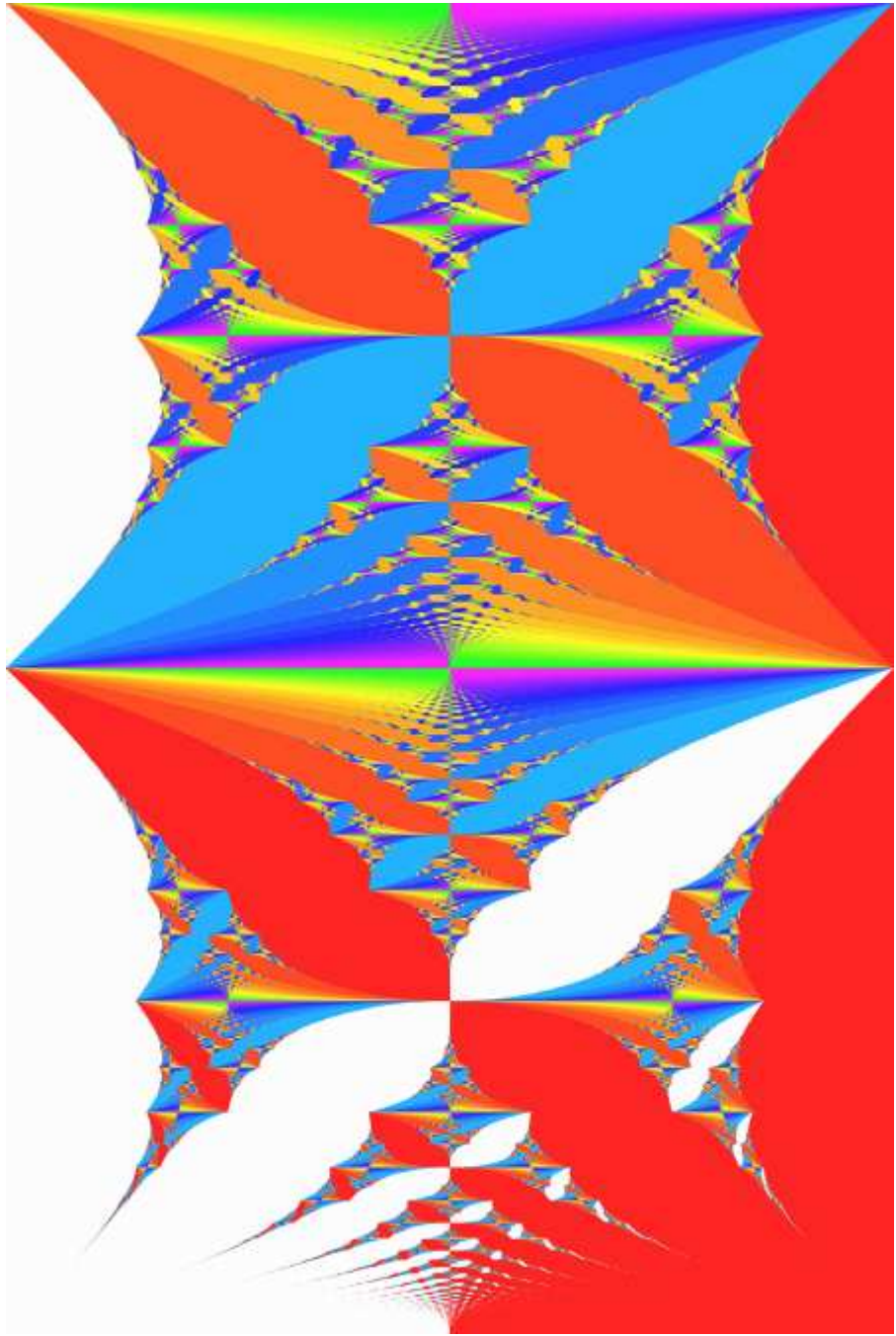} } 
\caption{Phase diagram of a split Landau level of a two dimensional electron gas
in a super-lattice. The  colors represent different
Chern numbers, or quantum Hall phases. The horizontal axis is chemical potential
which can be varied by changing the gate voltage, the vertical
axis is the number of lattice cells associated with a unit of quantum flux. } \label{hb-ll}
\end{figure}

The phase diagram of a split Landau level is shown in Fig. \ref{hb-ll}. The vertical axis is now {\em inversely} proportional to the magnetic field. That the two phase diagrams come from the same formal Hamiltonian is seen in the skeleton of the butterfly. However, once one pays attention to the colors, the two phase diagrams  are qualitatively different. The split Landau phase diagram  has less symmetry: It is is not periodic in $\phi$ and is not anti-symmetric under reflection.

Albrecht et.\ al.\ \cite{ref:albrecht} carried out experiments that test the predictions of the Hofstadter model of a split Landau level. The experiment successfully reconstructs the main features of the phase diagram of Fig.~\ref{hb-ll}.

According to conventional wisdom the integer Hall effect is a
feature of non-interacting electrons, while for strongly
interacting electrons fractions arise \cite{ref:jain}. One might
therefore expect that the theory of Chern numbers is restricted to non-interacting electrons. This is not so. Even with electron-electron interaction present quantum transport is related to Chern numbers: Consider an interacting 2D electron gas associated with a multiply connected system such as the one shown in Fig.~\ref{fig:riemann}. Thread the two holes with a pair of flux tubes.
The charge transport around one flux tube induced by the emf generated by the other flux tube is, again, related to adiabatic curvature. The adiabatic curvature is not quantized. However, its average  over the two fluxes is a Chern number and is therefore quantized.

\section{Fredholm indices}

An important theoretical model of the quantum Hall effect is that
of a 2D electron gas of independent electrons under the
influence of a random potential and strong magnetic field. Because
of the randomness there is no
Brillouin zone. (This is also the case for the Hofstadter model
with irrational magnetic field). Without a Brillouin zone to play the role of $S$ in Eq.~(\ref{eq:gauss}), the
Chern number approach to the Hall effect has no leg to stand on.

Jean Bellissard realized that  Hall conductance can be related to another
topological invariant, one that does not require a Brilloun zone:
The Fredholm index. It is a non-commutative analog of the Chern
number \cite{ref:bell}. Like the Chern number, the Fredholm index
is an integer which is stable under small deformations. In those
cases where the conductance may be interpreted either way the
Fredholm index coincides with the Chern number.

The Fredholm index is closely related to the creation and
annihilation operators of many-body quantum mechanics. The states
annihilated by an operator $F$ belong to a subspace of the Hilbert
space denoted by $ker F$. The dimension of this space, $\dim\, ker
F$, clearly an integer, counts the number of states annihilated by
$F$. It is tempting to relate the Hall conductance with the states
that were lost to reservoir $A$ upon the increase of $\Phi$ by a
quantum flux.

However, not every integer is a
topological invariant, and $\dim\, ker F$ is an example: It has no
stability.  If  $\dim\, ker F\neq 0$ then an arbitrarily small
deformation of $F$ will generically make $\dim\, ker F=0$.
It turns out that one can make a topological invariant
from  $\dim\, ker F$. This is the Fredholm index which is defined
by
\begin{equation}
Ind\ F =\dim\,ker F-\dim\ker F^*
\end{equation}
$Ind\ F$, being the  difference of dimensions, is an integer.  It
is  stable under deformations of $F$  because whenever $\dim\, ker
F$ jumps, so does  $\dim\, ker F^*$.

It can be shown  \cite{ref:bell,ref:ass} that the Hall conductance of a
two dimensional electron gas in the plane, in the presence of
random potential, is a Fredholm index, provided the Fermi energy
lies in the region of localized states. In fact,
Kubo's formula for the Hall conductance can be written as \cite{ref:ass}
\begin{equation}
\frac {e^2} h\,Ind\ (P_FUP_F)=\frac {e^2} h\,Tr\, \big(U[P_F,U^*]\big)^3
\end{equation}
Where $P_F$ projects on the states below the Fermi energy. $U$ is
a gauge transformation associated with one unit of quantum flux $
\Phi_0$.

How does this formula relate to the quantized staircase seen in
the Hall effect? The Fredholm index is not always defined: It may
be of the form $\infty-\infty$. However, when it is defined, it is
an integer. Once it is defined, it is stable under small
deformations. This implies that the Hall conductance as a function
of the chemical potential has quantized plateaus.

\section{Outlook}

The geometric and topological ideas developed in the context of the quantum Hall effect \cite{ref:thouless-niu} had substantial impact on an emerging new direction in mesoscopic physics:  Quantum pumps. These are  quantum dots that transport charge (or spin) between reservoirs that are otherwise in equilibrium. The pumps are modulated periodically by varying {\em two} independent parameters. For example, by manipulating the shape of the dot \cite{ref:marcus}.  For small pump cycle  the charge transport is proportional to the area and the constant of proportionality is an analog of the adiabatic curvature, as pointed out by Piet Brouwer \cite{ref:Brouwer}.

So far we have avoided saying anything about the Fractional quantum Hall effect. It is
time to redress this. The theory of the Fractional Hall effect \cite{ref:stone} is
based on the wave function approach of Robert Laughlin.  It developed into a large body of knowledge \cite{ref:stone,ref:jain}, some
of it, with strong geometric flavor. It is noteworthy that Chern numbers and Fredholm indices did not yet find their proper place in the theory of the fractional Hall effect. This is an open challenge.

Some of the ideas we touched upon here make contact also with
other areas of physics. Let us briefly note a relation to super-string theory and  quantum computing. In the former, there is much interest in
non-commutative geometry as a key to 
the structure of space time at the Planck scale.  The lowest
Landau level is the simplest  example of the non-commutative plane
and the Hofstadter model is a realization of the non-commutative
torus. In the latter, a major obstacle on the road to realizing quantum computers is to devise  qbits that are protected against decoherence and yet accessible for writing and reading. It has been proposed in \cite{ref:friedman-kitaev-larsen-wang} that the topological aspects of quantum Hall systems hold promise for realizing such qbits.

Paul Dirac said that {\em ``God used beautiful mathematics in creating the world''}. The story of topological quantum numbers in the Hall effect is an example.

\section{Acknowledgment} We thank F. Jegerlehner, U. Sivan and D. Thouless for useful comments.
This work is supported by the Technion fund for promotion of
research, by the EU grant HPRN-CT-2002-00277 and the Deutsche Forschungsgemeinschaft SFB288.

\end{document}